# Beyond Kinetic Harm and Towards a Dynamic Conceptualization of Cyberterrorism

VJ Straub[1]

[1]*Oxford Internet Institute*
*University of Oxford*
*Oxford, United Kingdom*

*E-mail: vincent.straub@oii.ox.ac.uk*

***Abstract*** *After more than two decades of discussion, the concept of cyberterrorism remains plagued by confusion. This article presents the result of an integrative review which maps the development of the term and situates the epistemic communities that have shaped the debate. After critically assessing existing accounts and highlighting the key ethical, social, and legal dimensions at stake in preventing cyberterrorist attacks, it calls for a more dynamic conceptualization that views cyberterrorism as more abstract, difficult to predict, and hard to isolate; and which embraces a different conception of sufficient harm. In concluding it proposes a novel definition of cyberterrorism, intended to catalyse a new research programme, and sketches a roadmap for further research.*

## 1. Introduction

As recent as 2018, 81% of Americans cited cyberterrorism as a leading threat to vital interests of the United States in the next 10 years (Norman 2018). Aside from capturing the public's attention, the term has also become a regular feature in the literature on cybersecurity and information warfare; yet, after more than two decades of discussion, its definition and dimensions still lack consensus and clarity (Macdonald *et al.* 2019). More significantly, whilst the National Academy of Sciences first warned of a 'digital Pearl Harbor' as early as 1990 (Weinberger 2013), most experts continue to agree that no identifiable instances of sophisticated 'pure cyberterrorism' have yet to materialize (Macdonald *et al.* 2019). The Stuxnet attack, first uncovered in 2010, is widely considered to be the only major external cyberattack of unknown origin known to have caused substantial kinetic (physical) damage including to the nuclear program of Iran (see, for example, Kenney 2015). Nonetheless, over the last ten years a shift in public perception has clearly occurred, coupled with the increasing sophistication and extent of cybercrime and revelations of mass government surveillance (Taylor *et al.* 2019). Next to ever-increasing Internet dependence and reliance on computer-mediated forms of interaction, political actors, industry, and civil society have started to pay more attention to cybersecurity, as the securitization of cyberspace has become a common talking-point in corporate boardrooms and the corridors of power. At the start of this decade, UN Secretary-General António Guterres proclaimed, "[we] must usher in order to the Wild West of cyberspace ... Cyberspace itself is at risk of cleaving in two" (Guterres 2020).

Against this backdrop, many scholars nevertheless continue to contend that the basic premise of the cyberterrorist threat has remained largely the same over time, that is, as societies have developed into mature information societies dependent on computer networks for their operation (Floridi 2012), new vulnerabilities are created—a massive 'electronic Achilles' heel' (Lewis 2002). Despite the seemingly uncontroversial nature of this claim, one of the central features of academic literature on cyber threats is the pace with which any single analysis becomes outdated. With this caveat in mind, a key aim of this article is to take stock of the various arguments that have been put forth over the last decades and provide a critical assessment of the current landscape of cyberterrorism discourse, particularly in lieu of the growing policy interest in identifying and pre-emptively counteracting cyber threats. Building on this, the second and main objective it is to shed light on some of the understudied social, legal, and ethical dimensions of counteracting cyberterrorism. In so doing, the key contribution of this article is to locate and situate novel interpretations of harm within recent conceptions of cyberterrorism, before providing an altogether new definition of cyberterrorism that integrates these perspectives and builds on findings emerging from cyberpsychology research as they relate to the difference between "non-physical" and kinetic damage. To conclude, this article provides recommendations for further inquiry.

Numerous studies have enriched our understanding of cyberterrorism over the last ten years or so. Key questions and topics that have been studied include: (i) the extent to which cyberterrorism is a real threat; (ii) whether known terrorist actors do or could conduct cyberterrorism; (iii) what the unique objectives behind an act of cyberterrorism might be; and (iv) the damage that can be caused by cyberterrorism (Conway 2018). Touching upon each one of these topics, the main thesis developed in this article builds on recent debates in information and cyber ethics (Allhoff 2017; Finlay 2018; Barbosa 2020) to argue that if practitioners and policymakers are to develop and implement adequate countermeasures to counter the potential of a cyberterrorist threat, a move beyond strict consequentialist interpretations of kinetic harm and a more dynamic conceptualization of cyberterrorism is required. Specifically, what is needed is a definition that explicitly considers cyberterrorism as more abstract, difficult to predict, and hard to isolate and which embraces a different conception of sufficient harm altogether—not constrained by an insistence on kinetic damage (Lucas 2017) and more attuned to forms of human suffering that may be less visible but still very real (Barbosa 2020). To advance this thesis and assess the state of the art, this article takes an integrative approach—it aims to build on a select number of influential articles from the academic (and grey) literature; rather than provide an overview of the entire subject (see Chen *et al.* 2014 for a comprehensive overview of the field; Veerasamy in Benson and McAlaney 2020 for a recent introduction), which is too wide to be covered in a single article (see Appendix A for details of the literature search method).

The structure of this article is as follows. Section 2 *Conflicting Perspectives in Conceptualizing Cyberterrorism* describes how cyberterrorist discourse has developed over time by synthesising literature from the fields of conflict studies, terrorist studies, cybersecurity, international relations, and computer security, among others. Thereafter, section 3 *Assumptions*

*and Implications in Responding to the Cyberterrorist Threat* identifies and analyses the validity of some of the key conceptualizations and framework that have been proposed. Section 4 *Discussion* examines some of the central social, legal, and ethical issues that remain unaddressed, especially as they pertain to state responsibility in counteracting the cyberterrorist threat. Finally, section 5 *A New Definition* considers recent debates in information and cyber ethics and situates non-kinetic interpretations of harm within recent conceptions of cyberterrorisms. To conclude, the article sketches a brief roadmap for future inquiry. In critically reviewing the concept of cyberterrorism and bringing together disparate lines of argument, this article will not only be of interest to scholars of cyberterrorism or those in related fields wishing to get up to speed on the current debate, it will also be useful to the increasing number of policymakers specialising or interested in understanding the nature and threat of cyberterrorism.

## 2. Conflicting Perspectives in Conceptualizing Cyberterrorism

It is a platitude to say that it is difficult to provide a specific and comprehensive account of cyberterrorism because the term terrorism is itself hard to define. Yet, whilst etymologists will tell us that the latter has unambiguous origins in the Latin t*errere* (meaning "to inspire fear" and first widely used during the French reign of terror to describe state-sponsored violence), it remains true that no single interpretation of 'terrorism' has yet gained broad acceptance (Williams 2009). In one sense, it should thus come as no surprise that when considering 'cyberterrorism' (cyber being an abbreviation of cybernetic, coined by Norbert Weiner and based on the Greek for "the one who governs"; *kybernetes*), the extent of definitional agreement does not increase. Quite the opposite, in fact. Since first entering discourse in the 1980s and early 1990s, when the spread of Information and Communication Technology (ICT) and arrival of the Internet sparked increased discussions on the potential risks faced by highly networked, technology-dependent societies, the term cyberterrorism has been variously cast. It has been described as perfectly capturing the 'sum of all fears' (Weimann 2005), an essentially contested concept (Weinberg *et al.* 2004), and even a 'red herring' (Clarke & Knake 2010). To unpack this contradictory situation, it helps to situate each view with respect to the different epistemic communities that have contributed to the term's evolution.

Initially employed as a general descriptor to refer broadly to novel acts of terrorism and crime committed through the use of hacking and computer resources (Parker 1983; Pollitt 1998), numerous refined definitions of cyberterrorism have been put forth over the last two decades. In perhaps the most influential version, formulated by computer scientist and information security researcher Denning (2001) at the turn of the millennium, the concept is taken to designate:

> "unlawful attacks and threats of attacks against computers, networks and the information stored therein when done to intimidate or coerce a government or its people in furtherance of political or social objectives. Further, to qualify as cyberterrorism, an attack should result in violence against persons or property …. Serious attacks against critical infrastructures could be acts of cyberterrorism, depending on their impact.

> Attacks that disrupt nonessential services or that are mainly a costly nuisance would not" (Denning 2001).

Informed by Denning and other early accounts (Lewis 2002; Valeri & Knights 2000), cyberterrorism has in turn been developed and adopted, at first cautiously within institutional settings by such organizations as the FBI and NATO (2008). Yet, upon closer inspection, it becomes clear that there are extensive discrepancies across various dimensions. Among others, differences exist when it comes to qualifying the exact objectives, targets, modes of operation, and significance of computer use. In the fields of conflict studies and computer security—which have tended to dominate the debate up until more recently—the biggest hurdle in arriving at consensual understanding arguably stems from the issue of whether to advance a narrow, 'targeted-oriented' view, or a broader, 'tool-oriented' one (Rizmal 2017). The narrow, 'pure cyberterrorist' definition, inspired by Denning (2001) and generally favoured by the security community tends to foreground the role of the perpetrators or 'agents' (often focusing on existing and known terrorist actors) and the physical damage they may cause to targets. By foregrounding the perpetrators' identities, such accounts have shed light on the sometimes-surprising motivational perspectives of cyberterrorist actors (Yunos and Sulaman 2017). Yet, in so doing this narrative also essentially maintains the portrayal of cyberterrorism as *acts by terrorists* using cyber-dependent means (Denning 2007; Gill *et al.* 2017; Jarvis & Macdonald 2014).

In contrast, broader conceptions have taken a multidimensional approach, stressing the novel role of cyberspace, that is, the online world distinct from everyday reality, and new technologies, and examining techniques, intent, motives, targets, and capabilities, alongside actors and effects (Veerasamy 2009; Plotnok and Sally 2019). Advocates of this approach have in turn also focused on investigating other terrorism-related activities besides attacks, such as online propaganda and recruitment (Chen *et al.* 2014; Conway 2017; Jarvis & Macdonald 2015). Recent illustrative examples oft cited have tended to include such cases as the web content created by so-called Islamic State (ISIS) (Giantas & Stergiou 2018). Kapsokoli (2019) provides a recent analysis of so-called ISIS's actions in social media to examine whether the transformative effect of cyberspace has changed the nature of terrorism, for instance.

One way in which further clarification has been sought, especially among scholars of conflict studies, is by distinguishing between cyberterrorism and other categories of cyberattacks that may nevertheless use similar methods, such as denial of service (DoS) attacks or malware (see Table 1). Here, the terms most often invoked for comparison include cybercrime, cyberwarfare, and hacktivism (Chen *et al.* 2014; Vegh 2002). Cybercrime, the broadest category, is usually taken to refer to the growing number crimes occurring each year which are exclusively or primarily motivated by financial or economic factors, for example, using social engineering and other forms of computer fraud (see Taylor *et al.* 2019), as opposed to political, social or otherwise ideological reasons (Roberts, 2018). Whilst cyberwarfare is used to describe cyberattacks that are directly enacted by governments or which are state-sponsored and targeted at other states (Janczewski & Colarik 2007; Lewis 2002); in contrast to cyberterrorist attacks, which are typically understood to be conducted by sub-national organizations or agents, understood to be conducted by sub-national organizations or agents, even operating within their own state. In this regard, the Stuxnet attack mentioned at the outset is now generally considered an instance of a state-sponsored cyberattack, rather than an act of cyberterrorism (see, for example, Kenney 2015 for full discussion). Finally, cyberterrorism has also come to be viewed as distinct from hacktivism. Although sometimes used loosely and interchangeably in early works (Denning 2001; Jordan and Taylor 2004), the latter is now chiefly employed to describe use of the Internet to achieve political or social objectives in a manner that is transgressive, nonviolent, and often inspired by civil disobedience, rather than intentions to coerce civilians or governments (Roberts 2018; Veerasamy 2020).

| Method of attack | Nature of attack |
| --- | --- |
| *Botnets* | After gaining access to a computer network and compromising all computers connected to the network, attackers can host automated software programs ("bots") that can conduct additional mass attacks. |
| *Denial of Service (DoS)* | DoS attacks employ computer code engineered to overwhelm or disrupt networks by reconfiguring network settings or impairing specific software applications by generating more traffic than can be handled. The use of a botnet to conduct large-scale DoS attacks is known as a Distributed Denial of Service (DDoS) |
| *Malware* | Malicious software ("malware") disrupts or comprises a computer system without the knowledge of the administrator or owner. After compromising a system, malware can be used for several purposes, such as reconfiguring settings or granting system access to the attacker |
| *Social Engineering* | In the context of a cyberattack, social engineering refers to the use of misleading digital communication to access information about or gain access to computer networks, systems, or users in order to then compromise, disrupt, or attack them (phishing is one such example whereby attackers pose as legitimate operators and may try to convince targets to provide personal data) |

**Table 1:** Descriptions of common methods of cyberattacks

A further approach taken by scholars who favour the broad definition has been to drop the cyberterrorism moniker altogether. Proponents of this persuasion argue that the examples of major cyberattacks like Stuxnet involve a level of technical expertise that far exceeds the current capacity of most (non-state) terrorist agents (Kenney 2015); and that the study of 'violent online political extremism' and 'covert political cyberaction' best describes the reality of terrorist intervention in cyberspace (Miller in Allhoff *et al.* 2016; Conway 2017; Gill *et al.* 2017). In other words, this line of argument suggests that it is important to focus on the methods and tactics by which the Internet and cyberspace have come to be primarily used by known terrorist organizations like the so-called ISIS.

Taken together, what is to be made of this lack of consensus and confusion, at times openly acknowledged in cyberterrorist discourse (Conway 2018; Stuart 2017)? Clearly the definitional issues around cyberterrorism remain as important to policymakers as they were since the term first entered the security lexicon, but it is evident that they far from being resolved. To unpack this issue, it is worth reflecting on the fact that since death or violence and damage that causes kinetic harm is often considered a necessary component of how cyberterrorism is conceived by most scholars (Jarvis & Macdonald 2015), much of this inconsistency is arguably a reflection of the fact that there have been few identifiable instances of sophisticated 'pure cyberterrorism' that have caused extensive harm (Macdonald *et al.* 2019). In other words, cyberterrorism remains in many ways more of a *potential* threat, making it difficult to analytically distil the issue in concrete terms. Indeed, the lack of real-world cyberterrorist catastrophes continues to be the reason why many scholars studying conventional terrorism and conflict have and continue to question the utility of terms like cyberterrorism and cyberwar altogether (Rid 2012a; Rid 2012b).

We can find additional support to bolster this line of reasoning if we consider the landscape of international law and norms applicable to cyberspace (Grove *et al.* 2000; Russell 2014; Leggat 2020). Although the remarks of Guterres mentioned in the introduction are somewhat hyperbolic, as cyberterrorism is by all accounts not a legal void—states have fundamental obligations to prevent, punish, and suppress malicious cyber conduct (Buchan 2016), for instance—it is widely accepted that cyberterrorist acts are not clearly encoded and thus not explicitly prohibited under international law. Rather, what constitutes as cyberterrorism still hinges on the country introducing the act (Maras 2016). Moreover, although many countries have laws which make reference to cyberterrorism (for example, the United Kingdom Terrorist Act of 2000 and the USA Terrorism Risk Insurance Act of 2002 extend to and have been amended to include cyberterrorism both in a narrow and broad sense; whilst other countries have introduced dedicated acts more recently, c.f. the Pakistani Prevention of Electronic Crimes Act of 2016 and the Kenya Computer Misuse and Cybercrimes Act of 2018), national and also regional laws often tend to simply treat cyberterrorism as any other act of terrorism (Kittichaisaree 2017; see also United Nations Office on Drugs and Crime 2020 for a list of relevant legislation). The stance first articulated by the European Union in 2013 is a good example of this. In its Cyber Strategy, the EU outlined that, "[t]he same laws and norms that apply in other areas of our day-to-day lives apply also in the cyber domain ... [the EU will] apply existing international laws in cyberspace" (European Union 2013). Yet, whilst this approach provides normative clarity, albeit mainly at a national level (Leggat 2020), it does not settle the issue concerning the specific way in which international law should be applied in cyberspace nor how we should understand the reality of cyberterrorist threats in light of continuing technological advancements (Brundage 2018).

In the face of definitional disagreement, some have argued that non-state cyberterrorism should simply be treated as a form of ordinary crime separate from international law including human rights law and the law of war (Marsili 2019). Whilst this may sound like a practical alternative for the short term, it downplays the risks and is clearly not sustainable in the long term, given the pace at which technology develops. In other words, progress on the legal level depends on the international community convincingly defining what constitutes acts of cyberterrorism.

In sum, the fact that detrimental cyberterrorist attacks as captured by the 'narrow approach' are still more a potential threat than a manifest reality let alone regular occurrence is arguably a central reason why cyberterrorism does not have a universally accepted definition, instead existing more as a consensus topic spanning a number of themes. These may in turn be captured as follows:
- for an attack to constitute an act of cyberterrorism, it must ultimately result in serious and intended *kinetic damage* in terms of human and economic casualties or intense fear among citizens;
- cyberterrorism is not random nor motivated purely by personal (financial) gain, rather it is *systematically* carried out by *non-state agents* inspired primarily by a political, social, religious, or ideological cause;
- to harm a government or a section of the public to varying degrees, cyberterrorism seriously interferes with critical *cyber-physical infrastructure* and *essential services* through the use *cyber-enabled means*.

Yet, as will be argued in the rest of this article, this general viewpoint, which has evidently not fundamentally changed in the last two decades, is perhaps no longer the most useful way to conceptualize cyberterrorism—if it is to be properly understood as a unique threat, qualitatively different from conventional terrorism. Rather, if practitioners and policymakers are to develop and implement adequate countermeasures in the coming decades, a more dynamic view may be required: one that explicitly considers cyberterrorism as more abstract, difficult to predict, and hard to isolate; and which embraces a different conception of sufficient harm, not constrained solely by the insistence on kinetic damage (Lucas 2017). Before developing this perspective and considering the practical recommendations it carries, it is instructive to first consider in the more depth the ways in which the various viewpoints sketched above have framed the social, legal, and ethical policy debate in responding to the threat of cyberterrorism, and consider which academic lacuna remain.

## 3. Assumptions and Implications in Responding to the Cyberterrorist Threat

Given the lack of definitional agreement and the fact that a significant number of lethal cyberterrorist attacks have yet to materialize, how have the central issues at stake in dealing with what many would view as a primarily theoretical threat been framed, particularly as they relate to the increasingly important questions of state responsibility in cyberspace (Taddeo 2020a)? To answer this question, this section (i) analyses the assumptions on which the various positions described in the previous section rest, and (ii) assesses the validity of the normative implications they carry—as these have ultimately come to shape discussion on governing cyberspace and responding to cyberterrorist threats (Christen *et al*, 2020; Lucas 2017).

### 3.1 Disregarding or underestimating the cyberterrorist threat

Let us first consider the position of those who are sceptical of cyberterrorism as well as cyberwar discourse in general and instead favour a recalibration of security efforts (Emery 2005; Rid 2012b). Whilst increasingly in the minority, a familiar premise invoked by scholars of this camp is the issue of non-falsifiability. In short, as cyberterrorism remains foremost a threat and no 'digital Pearl Harbor' has yet occurred, this is taken to simply underline how fortunate states have been; thereby ensuring the 'myth of cyberterrorism' (Emery 2005) lives on. Whilst seemingly reasonable, this logic nevertheless rests on shaky foundations. Aside the potential threat a lethal attack poses, a number of minor instances that may nonetheless be considered cyberterrorist in scope clearly have already occurred (see, for example, Center for Strategic and International Studies 2020 for a useful timeline of significant cyber events that have taken place since 2006). As a result, this view, call it the *present indicates the future* fallacy, which downplays the importance of needing to plan and implement unique cyberterrorist countermeasures, is not tenable in the medium to long term (aside from the aforementioned Stuxnet attack, it is worth emphasizing again that cybercrime and state-sponsored cyberattacks are already very much a current reality; see, for example, Ghafur *et al.* 2019 for a retroactive impact analysis of the 2017 WannaCry attack enacted by suspected North Korean state-backed hackers that infected over 300,000 computers in 150 countries causing billions of damage).

Turning to the position of scholars (concentrated in terrorism studies) who suggest we should shift our focus from the potential threat of a lethal cyberterrorist attack to the ways in which known terrorists use ICTs and the Internet for information warfare (Conway 2017; Smith 2015; Valeri & Knights 2000) we find other weaknesses. Whilst researchers belonging to this group convincingly employ empirical data to reveal current known terrorist use of ICT technologies "by the numbers" (Gill *et al.* 2017), the conclusion often reached, namely, that terrorists are likely to continue favouring the use of cyberspace for activities such as communication, propaganda, and intelligence gathering, is suspect for a number of reasons. Decreasing technical complexity, reduced costs and, perhaps most significantly, the ever-increasing development of digital technology and desire to link the virtual with the physical—consider the growth of autonomous vehicles and the proliferation of the Internet of Things—is creating new systems on a large scale vulnerable to serious cyberattacks (Greenberg 2015). This is coupled with the democratization and spread of cyber capabilities, which has become one of the most pronounced trends in recent years (Dorfman & Deppisch 2019). Custom malware that can be acquired using the Dark Web and freely available online hacking tutorials are just two simple examples; not to mention the potential malicious uses of machine learning and other artificial intelligence technologies (Brundage 2018). Thus, this *lack of capacity* view is fallacious in so far as it arguably not in tune with the growing technical reality. As the cybersecurity profession well knows, everything and anything can increasingly be hacked (Finch *et al.* 2013).

It is important to remember that an act like interrupting information sharing or turning off the traffic lights would bring ICT-dependent, 'hyper-historical' societies like the UK and others to a drastic halt (Floridi 2012); the Polish Tram Incident and the Los Angeles Traffic Surveillance Center Attack in the 2000s are two examples that provided small snapshots of this (see Maras

2015, p. 184). Although specific measures to tackle violent online political extremism, for instance, are undoubtedly necessary, it is arguably not enough to simply focus on known terrorist actors using ICTs: systematic security measures must be put in place that consider the interdependencies among known terrorists and the potential of lethal cyberterrorist threats. Without such measures in place, future cyberterrorist attacks may undermine the stability of information societies, as is the case with state development of artificial intelligence (Taddeo & Floridi 2018).

### 3.2 Securitization and the issue of disproportionate responses

Cyberterrorism is and must be treated as a significant threat, regardless of whether it materializes later rather than sooner or whether known terrorist agents are the perpetrators—this is the assumption we should start from (Conway 2018). A careful balance must nevertheless be struck in how we then decide to respond to this threat. Those who warn of 'cyberterrorism hype' should be applauded for warning us of the ways in which this threat can be used by political authorities to enact measures that may otherwise be taken to infringe upon civil liberties (Stuart 2017). That is, in lending our ear to those who warn of the grave threat cyberterrorism presents (Cronin 2019), we must be aware of the other side of the coin: the consequent 'securitization' of cyberspace (Aljunied 2019)—perhaps most prominently illustrated by citizen 'mass' surveillance (Palasinski & Bowman-Grieve 2017).

The theory of securitization helps alert us to the fact that national security, including cybersecurity, is not a natural given, rather, it can be carefully constructed by politicians and key decision-makers (McDonald 2008). As Roe (2008, p. 617) details, "in articulating a security threat, an actor thereby claims the special right to handle the issue using extraordinary means". In the wake of the ongoing global surveillance disclosures (Rogers & Eden 2017), the divide between what is and what is not security in a cyber context has become increasingly contentious, with broader conceptions of what can be threatened and how it can be threatened. The notion of 'state of exception' developed by Agamben (2005) in the wake of the 9/11 terrorist acts is useful here in distilling how governments may suspend laws and constitutional rights during a declared state of emergency that can in turn lead to a prolonged state of being, becoming the 'new normal'. At the time of writing, it is worth considering that over 60 countries have adopted some form of digital tracking and tracing system (DTTS) (Morley *et al.* 2020) to track the movements of their citizens as part of their response to mitigating the SARS-CoV-2 virus (in this case, a virus made of genetic material, as opposed to computer code). This would arguably have seemed farfetched only a few years earlier; whether these systems will all be scrapped by countries when the threat no longer warrants it remains an open question (Taddeo 2020b).

Importantly, cyberspace is also defined not just by its technical implementation, enabling communication to flow between networks of computers (and the surveillance of that communication), but by the social interactions involved (Madigan 2017). As such, a reoccurring problematic is the tendency to frame the threat of cyberterrorism as a double-barrelled question; 'cyberterrorism should be counteracted, and cyberspace securitized'. But accepting the threat of known terrorists using cyberspace and the potential threat 'pure' cyberterrorism poses does not mean we can forget the accountability of government officials. In asking how to adequately prepare for and counteract cyberterrorist attacks, it is imperative to consider how current measures, such as increased citizen surveillance, affect fundamental values like trust and privacy (Ford 2018). Respect for privacy—freedom from inappropriate interference or monitoring—is a stalwart of liberal democracies. Yet, the loss of privacy, reflected contemporaneously in such practices as the removal of information sharing between government agencies and use of biometrics, means any cyberterrorist countermeasures must not be indiscriminate, widespread, and politically convenient. Government securitization of critical cyber-physical infrastructure and essential online services cannot be equated with a clampdown on the Internet, for instance (Baldino & Lucas 2018). Thus, although the notion that "the privacy of a terrorist can never be more important than public safety" (Turnbull 2017) makes for a good political soundbite, it is arguably not a tenable political vision for a liberal democratic state, socially as well as ethically (Baldino & Lucas 2018).

## 4. Discussion

The central challenge in counteracting cyberterrorism is to balance the targeted mitigation of *existing* threats posed by known (cyber)terrorist agents using ICTs with a strategy that can adequately and proportionally protect mature information societies against the *potential* of a lethal cyberterrorist attack—whilst upholding liberal democratic values of trust and privacy. With regards to the latter, the previous section has sought to tease out some of the issues that arise when either an *inadequate* approach is taken (that is, leaving societies vulnerable) or a *disproportionate* policy response is adopted (for example, mass citizen surveillance). Focusing on the second of these two considerations, this section aims to identify avenues for further inquiry that can help to move the discussion forward, both conceptually and operationally, and prepares the groundwork for a novel interpretation of cyberterrorism, offered in the next section. To do so it considers the ways in which two common threads tacitly accepted in current debates may be impeding further progress. Specifically, building on prior work it investigates: (a) the centrality accorded to identifying (and prosecuting) the agents of cyberterrorism, and (b) the insistence on kinetic damage as a requisite for characterizing an attack as an act of cyberterrorism.

Although researchers are increasingly turning their attention to estimating the damage potential cyberterrorist attacks may pose (Evan *et al.* 2017), it is commonly accepted that cyberterrorist attacks are difficult to predict, given the variety of ways in which they may manifest (Chen *et al.* 2014). Some have suggested they be even not be understandable or commonsensical when compared to terrorist or non-terrorist attacks (Matusitz 2008). Consider the computer worm or virus, types of malware that may be weaponized by cyberterrorists to break into computer networks and infect them. Unlike conventional terrorism, this may have direct repercussions on a global scale; that is, it may result in a cascading failure, affecting all systems that are interconnected (He *et al.* 2016). This was the case with Stuxnet and the 2000 ILOVEYOU worm, albeit as an unintended consequence in the case of the latter (White 2020). It is also possible that security experts may not immediately detect the attack, let alone the attackers, nor know what exactly the virus has done. Further still, it is possible that after planting the virus, damage only occurs a week later and then stops, but then re-emerges several months later (Matusitz 2008). The crucial point here is simple: cyberterrorism may result in widespread disorder and chaos whilst all while it may remain impossible to easily isolate the attack, identify the perpetrators, or determine whether the threat has been mitigated.

This speaks to the broader issue at stake in an information age: our reliance on critical cyber-physical infrastructures and essential services may outstrip our capacity to protect them (Sullivant 2007). Similarly, the notion of finding (let alone prosecuting) the responsible agents ('vectors') may not easily translate to the realm of cyberterrorism (Roberts 2018). Even when known cyberterrorists hijack thousands of computer across the world, there are multiple issues involved simply in terms of prosecution; legal scholars have only recently begun to explore how extraterritorial extension under international law would need to operate in order for a country to declare a legitimate claim of prescriptive jurisdiction (Stockton & Golabek-Goldman 2013). Continuing this line of thinking it becomes clear that a more nuanced conception of cyberterrorism may be required; one which openly considers the threat as fundamentally more abstract, difficult to predict, and hard to isolate. The direct operational implication of such a standpoint would be to shift our strategic focus from active prevention via targeting identified agents of cyberterrorism to focusing exclusively on cybersecurity and shoring up the vulnerabilities in current cyber-physical systems (Taddeo *et al.* 2019).

To help identify these vulnerabilities, it is useful to return to the definitional issue of whether and to what extent cyberterrorism should only be labelled as such if it results in kinetic damage. Although scholars differ in the extent to which they characterize the nature of the physical harm that must be caused (that is, loss of life versus serious economic damage), the differences are arguably primarily matters of degree—as with discourse on cyberwarfare, what matters is that an attack causing some form of physical damage or harm has occurred (Lucas 2017). But what of cyberterrorist attacks that eventuate no physical damage (or result in no significant economic loss)? Of course, all activities in cyberspace are attacks on social interactions and physical people, as already mentioned. However, the criterion that a cyberterrorist attack must result in physical damage may perhaps have been too readily inherited from discourse on terrorism and thus be misguiding our attention to focus on only specific forms of infrastructure (that is, chemical reactor targets, transport infrastructure etc.; see, for example, Evan *et al.* 2017).

In stark contrast to prior assumptions, empirical studies are now also providing evidence to support this claim. Novel experimental findings from laboratory settings are beginning to suggest that cyberterrorists attacks carry significant psychological effects even in the absence of any physical harm (Gross *et al.* 2016; Gross *et al.* 2017). A recent study by Backhaus et al. (2020) exploring the emotional responses to cyberterrorism using specially designed news reports showing major cyber-attacks indicated that cyberterrorism arouses heightened reactions of anger and stress (measured physiologically through cortisol levels, and through self-report measures). Crucially, Backhaus et al. were able to show that (a) these emotional responses do not differ from the emotions triggered by conventional terrorism; and (b) were not dependent on the lethality of the attack. Additional observational research on how members of the public perceive and engage with cyber risk and how they are impacted during and after cyber-attacks has further illustrated the broad nature of the social and psychological aspects of such attacks (Bada and Nurse 2020). Whilst still in its infancy, this new line of research on cyberterrorist threats, cognitive vulnerabilities and cyberpsychology is already offering substantive evidence that cyber-attacks cause dire social disruption to people's daily lives, including anxiety or loss of confidence in cyber or technology; and significant psychological impact, including anxiety, worry, anger, outrage, depression, alongside stress. Whilst further follow-up work is required, such results offer tantalizing support for the claim that the effects of cyberterrorism are akin to conventional terrorism (and can even be worse, in some circumstances) even when its victims do not suffer injury or experience kinetic damage.

Alongside their empirical novelty, these findings also have far-reaching ethical consequences. In the context of cyberspace and cybersecurity governance, ethical and normative debates have so far tended to take consequentialist, deontological, and virtue ethics (Howard 2017) theories, among others, as their starting point. Consequentialism (consequence-based ethics), in particular, has had a strong influence on cyber discourse (Yaokumah 2020); which here refers to moral theories which hold that the consequences (i.e. outcomes) of a certain action or event form the foundation for any valid moral judgement about the specific action or create a framework for judgement (Loi and Christen 2020). This is evident in the continual insistence on what may be called 'kinetic end-states' as a qualifier of whether a cyber-attack may be deemed a cyberterrorist in nature (Blank 2015). To illustrate this, we must only take a simple example: if a cyberterrorist exposes somebody's Internet browsing history and online sexual activity to the world, it may cause social distress and psychological harm. This could in turn cause the victim emotional stress, anxiety, and depression, even if only temporary. Yet, from a strict consequentialist perspective, the severity and cyberterrorist nature of this attack ultimately depends on whether the psychological harm *leads* to material damages, such as divorce, loss of employment, suicide, or other so-called kinetic end-states. Whilst such an interpretation is helpful in clearly demarcating what should be counted as cyberterrorism, it arguably diminishes the "physical" non-visibility of harm and in turn perpetuates the lack of awareness around non-material forms of psychological suffering (Barbosa 2020).

Yet, an emerging and arguably more powerful interpretation of the aforementioned empirical findings on the harmful "non-physical" consequences of cyberattacks has recently offered by the likes of Allhoff (2017). In the context of cyberwarfare, he argues that the idea of 'physical' harm should be dropped altogether in favour of a more rule-utilitarian notion of 'sufficiently serious' harm. Extending this argument to the realm of cyberterrorism and building on the empirical findings described above, it can be argued that such an interpretation forces cybersecurity decision-makers to consider a much broader (but nonetheless important) range of vulnerabilities present in mature information societies that may be prone to cyberterrorist attacks. As an example, consider again not a lethal attack on a nuclear power plant or central banking system, but an attack aimed at taking down a major social media network site like Facebook, thereby potentially causing 'sufficient harm' to the lives of millions of citizens (Jenkins 2014). Do governments have a duty to their citizens to put security measures in place that explicitly aim to safeguard such sites from attack? In one sense, this view invites a significant shift in the dominant metaphysics of harm as found in the rulings of international law. Yet, in an altogether earthlier sense, it is also motivated by practical questions that warrant attention when the cyberterrorism threat is understood in the sense defined earlier and the growing evidence around the severe non-kinetic harm of cyberterrorism is taken into account. That is, it invites us to consider the possible need to establish a priori humanitarian regulations for mitigating the wide-ranging effects of potential cyberterrorist attacks. Importantly, whilst these new regulations, laws, and governance structures must necessarily be closely linked with the legal regulation of cyberterrorist actions, they should not be replaced or deprecated. As suggested by Barbosa (2020) in the context of cyberwarfare, any humanitarian regulations must be derived from a unique and altogether separate framework, grounded on the importance of safeguarding the humanitarian well-being of civilian populations affected by potential cyberterrorist attacks. In this respect, the 'sufficiently serious' principle provides a useful starting point for developing a more dynamic conceptualization of cyberterrorism that can help us develop such a framework.

## 5. A New Definition

There are fundamental and well-documented ethical issues at stake in countering cyberterrorism, notably in terms of citizen privacy, trust, and social cohesion. Over the last decade, many scholars working at the intersection of philosophy, international relations and security have shown a deep interest in simply applying the just war tradition to the *extra bellum* realm and questions of state responsibility (predominantly in the context of cyberwarfare), especially when it comes to "just surveillance" and "just intelligence" (Bellaby 2016; Macnish 2014; Quinlan 2007). Yet, as this article has sought to demonstrate, there remains a gap in analysing these topics when we begin to question dominant conceptions of cyberterrorism altogether, focus less on identifying the critical material elements of cyberterrorism, such as targets and agents (thus also calling into question dominants practices of mass surveillance as a means), and instead widen our understanding of how cyberterrorism may cause *harm*. This final section aims to open up this line of inquiry by providing an initial definition of cyberterrorism that incorporates the aforementioned findings emerging from cyberpsychology research and builds on the ethical points raised by the likes of Alhoff (2017), and other scholars of cyberethics (Floridi 2011), as they relate to the harm that can be caused by non-kinetic damage. It is hoped that this will inspire further work to consider the broader subsequent implications for current models of cybersecurity governance.

Currently, cyberterrorism is often still misconceived as cybercrime or confused with other forms of cyberattacks; to avoid such uncertainty, most scholars begin with definitions of terrorism and then formulate or adhere to definitions of cyberterrorism that entail distilling the term into its essential components. As discussed, these include 'actor' (e.g. non-state), 'motive' (e.g. ideological), 'means' (e.g. attack or threat of attack originating in cyberspace), 'target' (e.g. civilian) and so on (Veerasamy 2020). Most recently, Plotnek and Slay (2019, p. 3), for instance, offer a definition of cyberterrorism by first building a taxonomy of the key elements that feature in cyberterrorist discourse: "cyberterrorism is the premeditated attack or threat thereof by non-state actors with the intent to use cyberspace to cause real-world consequences in order to induce fear or coerce civilian, government, or non-government targets in pursuit of social or ideological objectives". Whilst they argue that their definition is meant to extend the list of critical elements typically included in prior definitions to also contain "[psychological] effect" (see ibid.), their insistence that cyberterrorism must result in "real-world consequences" nevertheless amplifies the importance of interpreting harm using consequentialist theories, thus falling prey to the drawbacks discussed earlier. Building on their definition and considering the arguments detailed throughout this article, it is in turn possible to construct a more universal definition of cyberterrorism as follows:

> "Cyberterrorism refers to attacks or threat thereof on information or on physical bodies and structures initiated by known and unknown non-state agents mediated by cyberspace that are intended to inflict destructive harm (primary effect), which may or may not be tied to real-world kinetic damage (secondary effect). To qualify as an act of cyberterrorism, the intended harm must compromise national security and or lead to, or have the potential to cause, civilian suffering, where such suffering is considered to be sufficiently serious regardless of whether it is the consequence of physical damage".

In distinguishing between primary and secondary impacts, this definition purposively draws inspiration from definitional frameworks adopted in the study of other threats, namely the assessment of natural disasters, such as earthquakes (Daniel 2017). This disaggregation helps to shift the focus away from actors and qualitative differences in harm (i.e. kinetic or, more specifically, corporeal, versus psychological) towards the magnitude of intended harm as the key variable against which to define any attack on civilian populations or governments as cyberterrorist in nature. The fact that cyberterrorist attacks must originate or occur in cyberspace in turn distinguishes cyberterrorism from conventional terrorism, whilst not drawing a hard border between the two.

To be sure, this definition stands in stark contrast to existing international law, as well as dominant consequentialist cyberethics. Yet, for reasons that need not be rehearsed here, the utilitarian basis on which this new definition is built is motivated by the desire to advance the discussion and help us build more secure information societies, focusing less on locating cyberterrorists and more on protecting citizens and their identities.

## 6. Conclusion

Although it remains foremost a theoretical threat, the potential of a 'pure' cyberterrorist attack and the grave consequences it would have is only growing. Yet, at the same time, a general issue within the study of cyberterrorism and relatedly cybersecurity is that both have not become widely recognized as cross-sector, multifaceted issues. In response, this integrative review has sought to illustrate that it is important to move across disciplinary boundaries, think beyond technology systems, image potential threats, and consider vulnerabilities as well as "non-physical" harms otherwise not immediately considered. The results of this analysis paved the way for the proposition of a new novel interpretation of cyberterrorism definition, introduced in section 5 *A New Definition,* which acknowledges the major contributions to the subject up until now and builds on recent empirical findings in cyberpsychology. By developing this definition and carrying on this line of inquiry into a broader ethical framework, we will be able to embed resilient cybersecurity systems and practices into the fabric of mature information societies.

Going forward, in the realm of international law states are evidently still at the beginning of the development of establishing norms of responsible behaviour in cyberspace. Although 2018 saw the launch of UN General Assembly Open-Ended Working Group (OEWG) and the Group of Governmental Experts (GGE), both dedicated to the issue (Stadnik 2019), they have not yet produced legally binding measures (United Nations 2020). However, the current national securitization of cyberspace is clearly not a sustainable approach. Further inquiry will thus benefit from considering the specific institutional means through which citizen oversight, alongside addressing the issues of privacy and trust, can be directly embedded into future cybersecurity measures. Fostering greater cybersecurity awareness in the next generation of computer science students, as recently suggested by Cruz and Simões (2020), is one very practical example that may be seen as a small step in the right direction, given the potential academic and career development paths such students may purse. But beyond this, the lack of effective cybersecurity measures and the potential knock-on effect this has on the current development of information societies around the globe (Floridi 2016) calls for more immediate macro action. In reconsidering the ethical frameworks underpinning the governance of cybersecurity, growing calls to focus on cyber peacekeeping (Papathanasaki *et al.* 2020) and to treat cybersecurity as a public good to be managed in the public interest (Mulligan and Schneider 2011; Weber 2017; Taddeo 2019) are in turn warmly welcomed as one further avenue of research that can help pave a path forwards towards a more secure future.

## Acknowledgments
The author would like to thank Mariarosaria Taddeo for providing the initial opportunity to conduct research on the topic of cyberterrorism at the Oxford Internet Institute and for providing useful comments, and William Silver for offering helpful input on an initial draft.

Gross, ML, Canetti, D, & Vashdi, DR, 2016, 'The psychological effects of cyber terrorism', *Bulletin of the Atomic Scientists*, vol. 72, no. 5, pp. 284–291.

Gross, ML, Canetti, D, & Vashdi, DR, 2017, 'Cyberterrorism: Its effects on psychological well-being, public confidence and political attitudes', *Journal of Cybersecurity*, vol 3., no. 1, pp. 49-58.

Grove, G, Goodman, S, & Lukasik, S, 2000, 'Cyber-attacks and international law', *Survival*, vol 42, no 3. pp 89–104.

Guterres, A, 2020, 'Remarks to the General Assembly on the Secretary-General's priorities for 2020', *United Nations*, viewed at 19 September 2020, <https://www.un.org/sg/en/content/sg/speeches/2020-01-22/remarks-general-assembly-priorities-for-2020>.

Hardy, K, & Williams, G, 'What is "Cyberterrorism"? Computer and Internet Technology in Legal Definitions of Terrorism' in *Cyberterrorism Understanding Assessment and Response.* New York: Springer. pp. 1–24.

He, Y, Smith, R, Janicke, H, Ayres, N, & Maglaras, LA, 2016, 'The mimetic virus: A vector for cyberterrorism', *International Journal of Business Continuity and Risk Management*, vol. 6, no. 4, pp. 259-271.

Howard, D, 2017, Civic Virtue and Cybersecurity in *The Nature of Peace and the Morality of Armed Conflict*, Demont-Biaggi F (Eds.), Palgrave Macmillan, Cham.

Hua, J, & Bapna, S, 2012, 'How Can We Deter Cyber Terrorism?', *Information Security Journal: A Global Perspective*, vol. 21, no. 2, pp. 102–114.

Janczewski, L, & Colarik, AM, 2007, *Cyber Warfare and Cyber Terrorism*, Hershey, New York, NY, US.

Jarvis, L, & Macdonald, S, 2014, 'Locating Cyberterrorism: How Terrorism Researchers Use and View the Cyber Lexicon', *Perspectives on Terrorism*, vol. 8, no. 2, pp. 52-65.

Jarvis, L, & Macdonald, 2015, 'What Is Cyberterrorism? Findings From a Survey of Researchers', *Terrorism and Political Violence*, vol. 27, no. 4, pp. 657–678.

Jarvis, L, Nouri, L, & Whiting, A, 2014, 'Understanding, Locating and Constructing Cyberterrorism' in Chen, TM, Jarvis, J, & Macdonald, S, (Eds.), *Cyberterrorism*, Springer, New York, NY, US.

Jenkins, FA, 2014, 'When Is a Real-World Response to a Cyberattack Justifiable? *Slate Magazine*, viewed at 20 September 2020,

Zotero, 2020, 'Zotero 5.0.', *Zotero*, viewed 20 September 2020, <https://www.zotero.org/>.

# Appendix A
# Procedure for Literature Review Search

This appendix describes how the integrative literature review (Torraco 2016) was executed. The first step was to identify the main objectives and key questions of interest. This was done by brainstorming for remaining gaps in conceptual understanding and describing the key ethical, legal, and social (ELS) dimensions of cyberterrorism with respect to traditional terrorism and non-terrorism, especially in terms of government responses and countermeasures. This resulted in differentiating between several interrelated thematic ELS issues, as discussed in the main text.

Second, these themes were used to collect relevant literature (illustrated in Figure A1). To begin, 3-5 "seeds" were identified for each of the themes to represent the most relevant articles. Exploratory keyword search was conducted for each theme using the Google search engine (www.google.com), with an intentional bias for impactful articles and inclusivity. Due to the emphasis on inclusivity, there was thus also a de-facto bias for older publications and grey literature. The content of the seed articles was then reviewed manually to compile lists of search keywords for each theme.

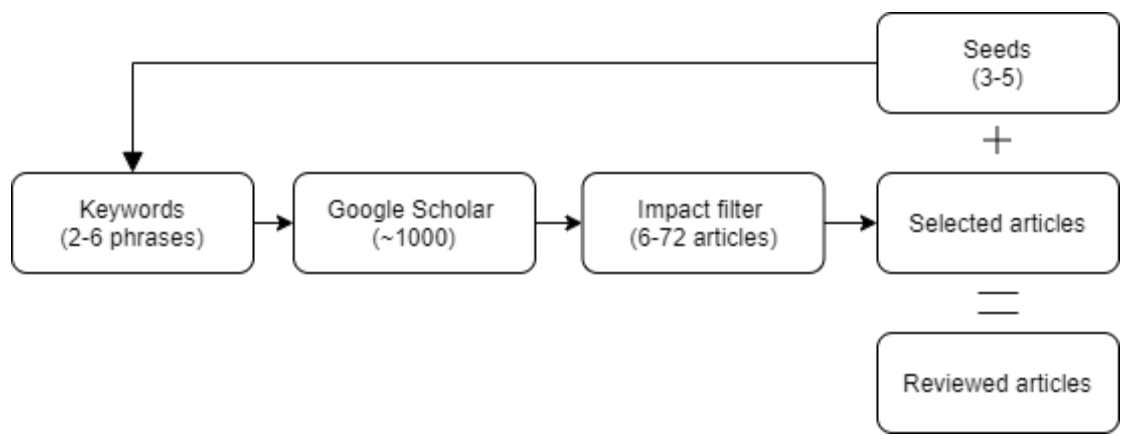

**Figure A1:** Method for the literature search

The lists of keywords were used for searches using Google and within the titles, abstracts, and keywords of all articles on the Google Scholar (www.scholar.google.com) and Semantic Scholar (www.semanticscholar.com) bibliographical databases. The first 1000 results were looked at, sorted, and, in the case of Google Scholar, the highest impact articles were selected, where impact relates to the number of citations. This method was taken to oversample and retrieve a diverse range of search results which can be grouped into a number of descriptive categories (see Table A1).

Finally, the research source management software Zotero (2020) and the Zotero Scholar Citation add-on (Beloglazov 2011) was used to sort items by citations. The abstracts were then read to manually identify the articles that focused on issues related to the each of the identified

ELS themes of cyberterrorism. This yielded between 6-10 articles (including the seeds) per thematic issue.